\documentclass[letterpaper, 10 pt, conference]{ieeeconf}  
\IEEEoverridecommandlockouts                              
\usepackage[utf8]{inputenc}
\usepackage[T1]{fontenc}
\usepackage{mathtools}
\usepackage[running]{lineno}

\usepackage{latexsym,amsmath,amssymb,amsfonts,graphicx}
\usepackage{epsfig}
\usepackage{setspace}
\usepackage{bm}

\newcommand*\patchAmsMathEnvironmentForLineno[1]{%
      \expandafter\let\csname old#1\expandafter\endcsname\csname #1\endcsname
      \expandafter\let\csname oldend#1\expandafter\endcsname\csname end#1\endcsname
      \renewenvironment{#1}%
         {\linenomath\csname old#1\endcsname}%
         {\csname oldend#1\endcsname\endlinenomath}}%
    \newcommand*\patchBothAmsMathEnvironmentsForLineno[1]{%
      \patchAmsMathEnvironmentForLineno{#1}%
      \patchAmsMathEnvironmentForLineno{#1*}}%
    \AtBeginDocument{%
    \patchBothAmsMathEnvironmentsForLineno{equation}%
    \patchBothAmsMathEnvironmentsForLineno{align}%
    \patchBothAmsMathEnvironmentsForLineno{flalign}%
    \patchBothAmsMathEnvironmentsForLineno{alignat}%
    \patchBothAmsMathEnvironmentsForLineno{gather}%
    \patchBothAmsMathEnvironmentsForLineno{multline}%
    }

\usepackage[textwidth=7in,textheight=9in,centering]{geometry}

\newtheorem{examplex}{Example}[section]
  {\popQED\endexamplex}
\usepackage{amssymb, amsmath, graphicx, enumerate}
\usepackage{csquotes}

\DeclareMathOperator*{\argmax}{\arg\!\max}

\usepackage{algorithm}
\usepackage[noend]{algpseudocode}
\newcommand{\bi}{\begin{itemize}}
\newcommand{\ei}{\end{itemize}}
\newcommand{\vo}[1]{\boldsymbol{#1}}

\newcommand{\X}{\vo{X}}

\usepackage{biblatex,xcolor}
\usepackage{empheq}
\usepackage[strict]{changepage}
\definecolor{darkgreen}{RGB}{0,128,0}

\usepackage[mode=buildnew]{standalone}
\usepackage{tikz}

\usepackage{verbatim}
\usepackage[most]{tcolorbox}
\allowdisplaybreaks

\title{Metrics for Bayesian Optimal Experiment Design under Model Misspecification}

\author{Tommie A. Catanach \& Niladri Das  
\thanks{
Tommie A. Catanach (tacatan@sandia.gov)  and Niladri Das (corresponding author, ndas@sandia.gov), are with Sandia National Laboratories, Livermore, CA 94550, USA.
 }}

\begin{document}
\maketitle
\begin{abstract}
The conventional approach to Bayesian decision-theoretic experiment design involves searching over possible experiments to select a design that maximizes the expected value of a specified utility function. The expectation is over the joint distribution of all unknown variables implied by the statistical model that will be used to analyze the collected data. The utility function defines the objective of the experiment where a common utility function is the information gain. This article introduces an expanded framework for this process, where we go beyond the traditional Expected Information Gain criteria and introduce the Expected General Information Gain which measures robustness to the model discrepancy and Expected Discriminatory Information as a criterion to quantify how well an experiment can detect model discrepancy. The functionality of the framework is showcased through its application to a scenario involving a linearized spring mass damper system and an F-16 model where the model discrepancy is taken into account while doing Bayesian optimal experiment design.
\end{abstract}

\section{Introduction}
For science and engineering systems there are often many choices of experiments to run or data to collect in order to infer information. Each of these choices has different costs in terms of time, money, or other constraints. One common solution to this problem stems from the field of Bayesian optimal experimental design (BOED). This approach uses the rigor of the Bayesian paradigm and information theory to formalize the design of experiments and treats it as an optimization problem. Concretely, the aim is to maximize a utility function that captures the worth of a particular experimental design. This utility function, typically the Expected Information Gain (EIG), depends on the posterior distribution sampled over many hypothetical realizations of plausible datasets from the experiment. However, for real applications, where there is the model discrepancy, EIG might not be the only relevant measure of information we should consider. 

In this work, we consider two additional criteria that measure notions of the robustness of  the design. The first criterion, Expected Generalized Information Gain (EGIG), captures the expected information gained (or lost) when an experimenter uses a model with discrepancy. The second criterion, Expected Discriminatory Information (EDI) reflects whether the information gained from an experiment would be sufficient to discriminate between the model and an alternative. The EGIG-based design seeks to mitigate discrepancy while the EDI-based seeks to only detect it. With these criteria, we aim to correct pathological issues in BOED and advance the BOED literature, which has a relatively few works concerning the robustness of BOED.

In \cite{brynjarsdottir2014learning} a Bayesian linear regression example is shown where the system is analysed without considering model discrepancies. There not only is the parameter under-estimated but the posterior credible intervals are not even close to
covering the true parameter value, which is alarming. In practice, despite the theoretical elegance and optimal performance for accurate models, BOED may encounter significant issues if our model is not properly specified. This means that there is no value of $\vo{x}^{\star}$ for which $p(\vo{y}|\vo{x} = \vo{x}^{\star}, \vo{d})$ corresponds to the true distribution for $p(\vo{y}|\vo{d})$, as noted in references \cite{Gr_nwald_2017}\cite{fengchi}. Although model misspecification is a common problem in Bayesian settings, BOED methods are especially vulnerable because they not only use the model to fit data, but also to generate new data. The main issue is that Bayesian approaches only account for uncertainty in the model parameters, not in the model's correctness, which can lead to disastrous outcomes where BOED continuously queries similar designs and produces low-quality datasets. Eliminating misspecification entirely is unrealistic, particularly in a general BOED context. However, there is still a lot of work that can be done to improve our comprehension and management of it. Presently, there is only a limited amount of research that covers both the theoretical \cite{farquhar2021statistical}\cite{fudenberg2017active} \cite{go2022robust}\cite{overstall2021bayesian}and empirical implications of misspecification \cite{sloman2022characterizing}, and very little has been done to examine the specific mechanisms that can lead to failures. This is where our EGIG and EDI metrics play an important role to evaluate the model robustness and identify modeling failures. Some Bayesian-adjacent approaches that call out the need for robustness and optimality in design are \cite{hernandez2019balancing} and \cite{pasquier2015robust}. Most notably, \cite{hernandez2019balancing} considers robust sensor placement for linear dynamical systems under asymptotic D-optimal design.

Outline: Section II introduces the model and key concepts, Sections III presents the BOED criteria, Section IV studies EGIG and EDI for two examples systems, and Section VII provides discussions.
\section{Modeling and Key Concepts}

\subsection{System Description}
Because of the difficulty of BOED, we will study this problem in the context of simplified models, specifically stationary discrete-time linear processes driven by Gaussian noise. We define the state vector as $\vo{x}_t\in \mathbb{R}^n$,
\begin{equation}\label{eq:lindyn}
    \vo{x}_t = \vo{A}\vo{x}_{t-1} + \vo{\eta}_t, \quad \quad (t=1,2,...).
\end{equation}
\noindent $\vo{A}$ is an $n \times n$ transition matrix and $\vo{\eta}_t \sim \mathcal{N}(\vo{0}, \vo{Q})$ is the process noise where $\vo{Q} \succeq 0$. We assume $\vo{x}_0 \sim \mathcal{N}(\vo{\mu}_0, \vo{\Sigma}_0)$. For simplicity, unless specified we will take $\vo{\mu_0} = 0$.  

The observation equation is
\begin{equation}\label{eq:linmeas}
    \vo{y}_t = \vo{H}\vo{x}_t + \vo{v}_t,
\end{equation}
where the measurements are $\vo{y}_t \in \mathbb{R}^s$, $\vo{H}$ is the measurement matrix and $\vo{v}_t \sim \mathcal{N}(\vo{0}, \vo{R})$ where
$\vo{R} \succ \vo{0}$. The random vectors $\{\vo{x}_0, \vo{\eta}_1,..., \vo{\eta}_t, \vo{v}_1,..., \vo{v}_t\}$ are assumed to all be independent.

From this general case, we will study two simplifications. First, we consider a system without dynamics (or equivalently a single time step of the system), corresponding: $\vo{y} = \vo{H}\vo{x} + \vo{v}$. Second, we will study the system after it has converged to its stationary distribution, assuming that $\vo{A}$ is asymptotically stable. In this case, if $t$ is sufficiently large, we have that $\vo{x}_t \sim \mathcal{N}(\vo{0}, \vo{\Sigma}_L)$, where $\vo{\Sigma_L}$ is the solution to the discrete Lyapunov equation  $\vo{\Sigma_L} = \vo{A}\vo{\Sigma_L}\vo{A^T} + \vo{Q}$.
\subsection{Bayesian Inference}
In Bayesian inference to rigorously update our beliefs about $\vo{X}$ with observation data $\vo{Y}$, we apply Bayes' theorem,
\begin{equation}
    p \left (\vo{X} \mid \vo{Y} \right) = \frac{p\left ( \vo{Y} \mid \vo{X} \right)p \left (\vo{X} \right)}{p\left ( \vo{Y} \right)}.
\end{equation}
The prior $p \left (\vo{X} \right)$ reflects our initial beliefs about $\vo{X}$ while $p \left (\vo{X} \mid \vo{Y} \right)$ is our posterior (after observations) belief. The likelihood, $p\left ( \vo{Y} \mid \vo{X} \right)$ is the probability of observing $\vo{Y}$ given a state $\vo{X}$, while $p\left ( \vo{Y} \right)$ is the overall probability of observing the data given our prior (called the evidence). Often we are interested in measuring how informative is the data. To do this we measure our change in belief, i.e. the information gain, using the Kullback–Leibler (KL) divergence,
\begin{equation}
\text{D}_{\text{KL}} \left [ p \left (\vo{X} \mid \vo{Y} \right) || p \left (\vo{X} \right) \right] = \int p \left (\vo{X} \mid \vo{Y} \right) \log \frac{p \left (\vo{X} \mid \vo{Y} \right)}{p \left (\vo{X} \right)} d\vo{X}
\label{eq:KL_def}
\end{equation}
For the Gaussian case where $p \left (\vo{X} \mid \vo{Y} \right) \sim \mathcal{N} \left (\vo{\mu_1}, \vo{\Sigma_1} \right )$ and $p \left (\vo{X} \right) \sim \mathcal{N} \left (\vo{\mu_0}, \vo{\Sigma_0} \right )$ the KL divergence is
{\footnotesize
\begin{align}
    \frac{1}{2} (\text{Tr} \left [\vo{\Sigma_0}^{-1} \vo{\Sigma_1} \right] - n + \left ( \vo{\mu_1} - \vo{\mu_0} \right )^T \vo{\Sigma_0}^{-1} \left ( \vo{\mu_1} - \vo{\mu_0} \right )
    + \log \frac{\mid\vo{\Sigma_0}\mid}{\mid\vo{\Sigma_1}\mid} ).
    \label{eq:KL_MVN}
\end{align}}

The KL divergence can be generalized using a more expressive, yet still information theoretically valid, measure of information \cite{e22010108} defined over three distributions: $r(\vo{X})$, $p(\vo{X})$, and $q(\vo{X})$ given by
\begin{equation}
\mathcal{I}_{r(\vo{X})} [p(\vo{X}) \mid \mid q(\vo{X})] = \int r \left (\vo{X}\right) \log \frac{p \left (\vo{X} \right)}{q \left (\vo{X} \right)} d\vo{X}.
\label{eq:gen_info}
\end{equation}
The interpretation of this form of information is that we want to measure a change in belief (e.g. information gained or lost) when updating from   
$q \left (\vo{X} \right)$ to $p \left (\vo{X} \right)$ in the view of $r \left (\vo{X}\right)$. The view defines our reference frame for assessing changes in information. Typically, both $r \left (\vo{X}\right)$ and $p \left (\vo{X} \right)$ would be the posterior and $q \left (\vo{X} \right)$ the prior, recovering the KL divergence. However, in the case where there is a model discrepancy, $r \left (\vo{X}\right)$ could be the unknown posterior from the true model, while $p \left (\vo{X} \right)$ could the inferred posterior from the model with discrepancy. Therefore, we could measure whether inference with the model discrepancy is still getting close to the correct result. We note that unlike the KL divergence this measure can be negative meaning that $q \left (\vo{X} \right)$ provides more information about $r \left (\vo{X}\right)$ than $p \left (\vo{X} \right)$ does.

For the case where, $r(\vo{X})$, $p(\vo{X})$, and $q(\vo{X})$ are all described by multivariate Gaussians,
\begin{align}
&\mathcal{I}_{r(\vo{X})} [p(\vo{X}) \mid \mid q(\vo{X})] = \nonumber \\ &\frac{1}{2} \bigg (\text{Tr} [(\vo{\Sigma_q}^{-1}-\vo{\Sigma_p}^{-1}) \vo{\Sigma_r}]  -\left ( \vo{\mu_r} - \vo{\mu_p} \right )^T \vo{\Sigma_p}^{-1} \left ( \vo{\mu_r} - \vo{\mu_p} \right ) \nonumber \\
&+\left ( \vo{\mu_r} - \vo{\mu_q} \right )^T \vo{\Sigma_q}^{-1} \left ( \vo{\mu_r} - \vo{\mu_q} \right ) + \log \frac{\mid\vo{\Sigma_q}\mid}{\mid\vo{\Sigma_p}\mid} \bigg).
\label{eq:gen_info_gauss}
\end{align}

\noindent This uses the fact that Eq. \ref{eq:gen_info} can be expressed as the difference of two KL divergences and employing Eq. \ref{eq:KL_MVN}.

\subsection{Bayesian Filtering}
For a Markov process where the state $\vo{x_t}$ only depends on $\vo{x_{t-1}}$ and the observation $\vo{y_t}$ only depends on $\vo{x_t}$ we can simplify the inference problem for the state $\vo{x_t}$ given a time series of observations $\vo{Y_t}=\{\vo{y_0} \dots \vo{y_t} \}$ as 

\begin{equation}
     p \left (\vo{x_t} \mid \vo{Y_t} \right) = \frac{p\left ( \vo{y_t} \mid \vo{x_t} \right)p \left (\vo{x_t} \mid \vo{Y_{t-1}} \right)}{p\left ( \vo{Y} \mid \vo{Y_{t-1}} \right)}.
\end{equation}

Using this, the Bayesian filter for the system described by Eq.(\ref{eq:lindyn})-(\ref{eq:linmeas}), is the Kalman filter,
\begin{align}
\label{eq:kalman_filter_first}
    \vo{\mu}_{t|t-1} &= \vo{A}\vo{\mu}_{t-1|t-1} \\
    \vo{\Sigma}_{t|t-1} &= \vo{A}\vo{\Sigma}_{t-1|t-1} \vo{A}^T + \vo{Q}\\
    \vo{\mu}_{t|t} &= \vo{\mu}_{t|t-1} + \vo{K}_t(\vo{y}_t-\vo{H}\vo{\mu}_{t|t-1} )\\
    \vo{\Sigma}_{t|t} &= (\vo{I}-\vo{K}_t\vo{H})\vo{\Sigma}_{t|t-1}\label{eq:kalman_filter_last}.
\end{align}
where $\vo{K}_t = \vo{\Sigma}_{t|t-1}\vo{H}^T\vo{S_t}^{-1}$ is the Kalman gain matrix and $\vo{S_t} = \vo{H}\vo{\Sigma}_{t|t-1}\vo{H}^T + \vo{R}$ is the predictive uncertainty.
Considering a single time step, the \textit{a-priori} estimator of $\vo{x}_t$ is $\vo{\mu}_{t|t-1}$ with covariance $\vo{\Sigma}_{t|t-1}$. The \textit{a-posteriori} estimator of $\vo{x}_t$ is $\vo{\mu}_{t|t}$ with covariance $\vo{\Sigma}_{t|t}$. Therefore, the prior, posterior, and evidence are
\begin{align}
    p(\vo{x}_t) &\sim \mathcal{N}(\vo{\mu}_{t|t-1} ,\vo{\Sigma}_{t|t-1}),\\
    p(\vo{x}_t  \ | \ \vo{y}_t,\vo{d}) &\sim \mathcal{N}(\vo{\mu}_{t|t} ,\vo{\Sigma}_{t|t} ),\\
    p(\vo{y}_t \ | \ \vo{d}) &\sim \mathcal{N}(\vo{H}\vo{\mu}_{t|t-1},\vo{S_t}).
    \label{eq:kal_pred}
\end{align}

As we can see from Eq. \ref{eq:kalman_filter_first} - \ref{eq:kalman_filter_last}, only the means $\vo{\mu}$ depend on the observations $\vo{y}$. Thus, when $\vo{A}$ is asymptotically stable we can find the stationary distribution of $\vo{\Sigma}_{t|t}$. Here we define $\vo{\Sigma}_{t|t-1} \rightarrow \vo{\Gamma}$ and $\vo{\Sigma}_{t|t} \rightarrow \vo{\Sigma}_D$ as $t \rightarrow \infty$.
We do this by first using the discrete time algebraic Riccati equation (DARE) given by
\begin{align}
&\vo{\Gamma} = \vo{A} \vo{\Gamma} \vo{A}^T + \vo{Q} - \vo{A} \vo{\Gamma} \vo{H}^T (\vo{H} \vo{\Gamma} \vo{H}^T + \vo{R} )^{-1} \vo{H} \vo{\Gamma} \vo{A}^T
\label{eq:DARE1}
\end{align}
and then solve for $\vo{\Sigma}_D$ via
\begin{equation}
\vo{\Sigma}_D = \vo{\Gamma}-\vo{\Gamma}\vo{H}^T(\vo{H}\vo{\Gamma}\vo{H}^T + \vo{R})^{-1}\vo{H}\vo{\Gamma}.
\label{eq:DARE2}
\end{equation}
\subsection{Bayesian Optimal Experimental Design}
In BOED, the first step to modeling the problem is to define a utility function $U(\vo{d})$ that gives the value of performing an experiment at $\vo{d}\in\mathcal{D}$. The set $\vo{d}\in\mathcal{D}$ spans the space of possible designs. In Bayesian design, the utility is a function of the posterior distribution $p(\vo{X} \ | \ \vo{d},\vo{Y})$. The utility function is maximized to find the optimal design $\vo{d}^*$, i.e. $\vo{d}^* = \argmax_{\vo{d}\in\mathcal{D}} U(\vo{d})$.

The choice of the utility function $U(\vo{d})$ is crucial, as different functions will usually lead to different optimal designs\cite{ginebra2007measure}. One of the most principled choices often used in BOED is the mutual information. This is the information gained about $\vo{X}$ by taking measurements, $\vo{Y}$, according to design $\vo{d}$. This is just the KL divergence from the prior to posterior, $\text{D}_{\text{KL}}[p(\vo{X}  \ | \ \vo{Y},\vo{d})|| p(\vo{X})]$, Eq. \ref{eq:KL_def}.

However at the point of choosing $\vo{d}$, we do not have the measurements. Thus, in order to access the effectiveness of the design $\vo{d}$, we take the expected KL divergence over plausible data sets $p(\vo{Y} \ | \vo{d})$. This utility function is known as the Expected Information Gain (EIG) and is defined as,
\begin{align}
\text{EIG}(\vo{d}) &= \mathbb{E}_{p(\vo{Y} \ | \vo{d})}[\text{D}_{\text{KL}}[p(\vo{X}  \ | \ \vo{Y},\vo{d})|| p(\vo{X})] ] \nonumber \\
&= \int p(\vo{X}, \vo{Y} \ | \ \vo{d}) \log \frac{p(\vo{X}  \ | \ \vo{Y},\vo{d})}{p(\vo{X})} d\vo{X} d\vo{Y}.
\end{align}

\section{Bayesian Optimal Experimental Design Criteria}
\subsection{Expected Information Gain}

For the linear Gaussian model given by Eq.(\ref{eq:lindyn})-(\ref{eq:linmeas}), we can derive expressions for the EIG.

\noindent \underline{\emph{Single Step Update}}: First for the case of a single update step (or equivalently when no dynamics are present) we begin by substituting the values from Eq. \ref{eq:kalman_filter_first} - \ref{eq:kalman_filter_last} into the Gaussian KL divergence expression, Eq. \ref{eq:KL_MVN}. Rearranging terms with the matrix inversion lemma and cyclic property of the trace, the information gain from the prior to the posterior is
\begin{align}
    &\text{D}_{\text{KL}}(p(\vo{x}_t  \ | \ \vo{y}_t,\vo{d})|| p(\vo{x}_t)) =\frac{1}{2}\Big[\text{log}|\vo{I} +\vo{H}^T\vo{R}^{-1}\vo{H}\vo{\Sigma}_{t|t-1}|\nonumber\\ 
    & - \text{tr}[\vo{S_t}^{-1}\vo{H}\vo{\Sigma}_{t|t-1}\vo{H}^T]\nonumber\\
    &+(\vo{y}_t-\vo{H}\vo{\mu}_{t|t-1})^T\vo{S_t}^{-1}\vo{H}\vo{\Sigma}_{t|t-1}\vo{H}^T\vo{S_t}^{-1}(\vo{y}_t-\vo{H}\vo{\mu}_{t|t-1})\Big]
    \label{eq:KL_KD}
\end{align}
\noindent Only the last term depends on $\vo{y}_t$, so for the EIG we just need to find the expectation of the quadratic term, which is
\begin{align}
    &\mathbb{E}_{p(\vo{y}_t | \vo{d})} [ (\vo{y}_t-\vo{H}\vo{\mu}_{t|t-1})^T\vo{S_t}^{-1}\vo{H}\vo{\Sigma}_{t|t-1}\vo{H}^T\vo{S_t}^{-1}(\vo{y}_t-\nonumber \\& \quad \vo{H}\vo{\mu}_{t|t-1})] \nonumber \\
    &= \text{Tr} [\vo{S_t}^{-1}\vo{H}\vo{\Sigma}_{t|t-1}\vo{H}^T\vo{S_t}^{-1} \text{Cov} (\vo{y}_t-\vo{H}\vo{\mu}_{t|t-1}) ] \nonumber\\
    &= \text{Tr} [\vo{S_t}^{-1}\vo{H}\vo{\Sigma}_{t|t-1}\vo{H}^T].
    \label{eq:E_quad}
\end{align}
Here we recall Eq. \ref{eq:kal_pred} so $\vo{y}_t-\vo{H}\vo{\mu}_{t|t-1}$ has mean $\vo{0}$ and covariance $\vo{S}_t$. Therefore, noting the cancellation of the trace terms, the EIG of the single step of the Kalman filter is
\begin{tcolorbox}[ams align]
\text{EIG}(\vo{d}) &= \mathbb{E}_{p(\vo{y}_t \ | \ \vo{d})}[\text{D}_{\text{KL}}(p(\vo{x}_t  \ | \ \vo{y}_t,\vo{d})|| p(\vo{x}_t))]\nonumber\\
&= \frac{1}{2} \log \frac{| \vo{\Sigma}_{t|t-1} |}{| \vo{\Sigma}_{t|t} |} \nonumber \\
&=\frac{1}{2}\Big[\text{log}|\vo{I}+\vo{H}^T\vo{R}^{-1}\vo{H}\vo{\Sigma}_{t|t-1}|\Big].
\label{eq:expectutil}
\end{tcolorbox}

\underline{\emph{Infinite Horizon}}: We may also be interested in assessing the EIG about a state $\vo{x}_t$ when the system and filters have converged to their stationary distributions. For this we define our prior knowledge about $\vo{x}_t$ as solution to the Lyapunov equation e.g. $p(\vo{x}_t | \vo{d}) = \mathcal{N}(\vo{0}, \vo{\Sigma}_L)$ when $t$ is  sufficiently large to be in the asymptotic regime. Similarly, when we have a sufficiently large set of observations, $\vo{Y}_t$, we know the posterior belief about $\vo{x}_t$ will have the form $p(\vo{x}_t | \vo{Y}_t, \vo{d}) = \mathcal{N}(\vo{\mu}_t(\vo{Y}_t), \vo{\Sigma}_D)$. Here we express $\vo{\mu}_t$ as a function of $\vo{Y}_t$ to emphasize that $\vo{\mu}_t$ is a random variable defined by $\vo{Y}_t$. Therefore, the information gain from observing the $\vo{Y}_t$ is
\begin{align}
\label{eq:KL_Asymp}
&\text{D}_{\text{KL}}(p(\vo{x}_t | \vo{Y}_t, \vo{d})|| p(\vo{x}_t)) = \\
&\frac{1}{2} \left ( \text{Tr} \left [\vo{\Sigma}_L^{-1} \vo{\Sigma}_D \right] - n +  \vo{\mu}_t(\vo{Y}_t)^T \vo{\Sigma}_L^{-1} \vo{\mu}_t(\vo{Y}_t) + \log \frac{\mid \vo{\Sigma}_L \mid}{\mid\vo{\Sigma}_D\mid} \right ).\nonumber 
\end{align}
Again, the only term that depends on the observations is the quadratic term. Therefore, to compute the EIG we first derive the expectation,
\begin{align}
    \mathbb{E}_{p(\vo{Y}_t|\vo{d})} &[\vo{\mu}_t(\vo{Y}_t)^T \vo{\Sigma}_L^{-1} \vo{\mu}_t(\vo{Y}_t) ] \nonumber \\
    &= \text{Tr} [\vo{\Sigma}_L^{-1} \text{Cov} (\vo{\mu}_t(\vo{Y}_t) \vo{\mu}_t(\vo{Y}_t)^T) ] \nonumber \\
    &= \text{Tr} [\vo{\Sigma}_L^{-1} (\vo{\Sigma}_L - \vo{\Sigma}_D) ] = n - \text{Tr} [\vo{\Sigma}_L^{-1} \vo{\Sigma}_D ].
    \label{eq:Asmp_E_quad}
\end{align}
Here we use that $\mathbb{E}_{p(\vo{Y}_t|\vo{d})} [\vo{\mu}_t(\vo{Y}_t)] = \vo{0}$ and $\mathbb{E} [ \vo{\mu}_t(\vo{Y}_t) \vo{\mu}_t(\vo{Y}_t)^T] = \vo{\Sigma}_L - \vo{\Sigma}_D$. This is shown as Eq. \ref{eq:mu_cov} in Appendix \ref{sec:asymp_means}.

Therefore, taking the expectation of Eq. \ref{eq:KL_Asymp} over $\vo{Y}_t$ and substituting in the result of Eq. \ref{eq:Asmp_E_quad} which cancels the trace terms, we find similarly to Eq. \ref{eq:expectutil} that

\begin{tcolorbox}[ams align]
\text{EIG}(\vo{d}) &= \mathbb{E}_{p(\vo{Y}_t \ | \ \vo{d})}[\text{D}_{\text{KL}}(p(\vo{x}_t  \ | \ \vo{Y}_t,\vo{d})|| p(\vo{x}_t))]\nonumber\\
&\rightarrow \frac{1}{2} \log \frac{| \vo{\Sigma}_{L} |}{| \vo{\Sigma}_{D} |}, \text{ as } t \rightarrow \infty.
\label{eq:asymp_EIG}
\end{tcolorbox}

\subsection{Expected Generalized Information Gain}
Using the generalized measure of information in eq. \ref{eq:gen_info}, we can assess how much information is expected to be gained or lost by an experiment $d$ when there is a model discrepancy. We define the true model as $\mathcal{M}^*$ and the model with discrepancy as $\mathcal{M}$, both of which have the same unknown states, $\vo{X}$ which we seek to infer. This expectation is taken over data that is generated according to $p(\vo{Y} \ | \vo{d}, \mathcal{M}^*)$. This leads to the Expected Generalized Information Gain (EGIG) given by,
\begin{align}
&\text{EGIG}(\vo{d}, \mathcal{M}, \mathcal{M}^*) = \nonumber\\
& \mathbb{E}_{p(\vo{Y} \ | \vo{d}, \mathcal{M}^*)} \bigg[\mathcal{I}_{p(\vo{X}  \ | \ \vo{Y},\vo{d},\mathcal{M}^*)}[p(\vo{X}  \ | \ \vo{Y},\vo{d}, \mathcal{M})|| p(\vo{X} \mid \mathcal{M})] \bigg ] \nonumber \\
&= \int p(\vo{X}, \vo{Y} \ | \ \vo{d}, \mathcal{M}^*) \log \frac{p(\vo{X}  \ | \ \vo{Y},\vo{d}, \mathcal{M})}{p(\vo{X} \mid \mathcal{M})} d\vo{X} d\vo{Y}\\
&= \int p(\vo{X}, \vo{Y} \ | \ \vo{d}, \mathcal{M}^*) \log \frac{p(\vo{Y}  \ | \ \vo{X},\vo{d}, \mathcal{M})}{p(\vo{Y} \mid \mathcal{M})} d\vo{X} d\vo{Y}
\label{eq:EGIG_Evidence}
\end{align}
\noindent Notes that Eq. \ref{eq:EGIG_Evidence} is a simple rearrangement using Bayes' theorem, which can be easier to compute for some problems. 

Typically we do not know the model $\mathcal{M}^*$, so in practice we should either define a set of plausible models we want to be robust to or we can assess the sensitivity to perturbations away from $\mathcal{M}$ by computing derivatives of the EGIG using either automatic differentiation or numerical derivatives.

In the context of inferring $\vo{x}_t$ with a system defined by Eq. \ref{eq:lindyn} - \ref{eq:linmeas} we define the true model $\mathcal{M}^* = \{\vo{A}^*, \vo{H}^*, \vo{Q}^*, \vo{R}^*\}$ and the model we use for inference as $\mathcal{M} = \{\vo{A}, \vo{H}, \vo{Q}, \vo{R}\}$.\\

\noindent \underline{\emph{Single Step Update}}: We start with the EGIG form of Eq. $\ref{eq:EGIG_Evidence}$. We defined $\vo{\mu}_{t|t-1} = \vo{A}$, $\vo{\mu}_{t|t-1}^* = \vo{A}^*$, $\vo{\Sigma}_{t|t-1} = \vo{A}\vo{\Sigma}_{t-1|t-1} \vo{A}^T + \vo{Q}$, and $\vo{\Sigma}_{t|t-1}^* = \vo{A}^*\vo{\Sigma}_{t-1|t-1}^* \vo{A}^{*T} + \vo{Q}^*$. We then note the distributions,
\begin{align}
&p(\vo{x}_t,  \vo{y}_t \mid d, \mathcal{M}^*) =\label{eq:xy_joint}\\
&\mathcal{N} \Bigg( \begin{pmatrix}
\vo{\mu}_{t|t-1}^* \\
\vo{H}^*\vo{\mu}_{t|t-1}^*
\end{pmatrix},
\begin{pmatrix}
\vo{\Sigma}_{t|t-1}^*  & \vo{\Sigma}_{t|t-1}^*\vo{H}^{*T} \\
\vo{H}^{*}\vo{\Sigma}_{t|t-1}^* & \vo{S}_t^*
\end{pmatrix}\Bigg)\nonumber
\end{align}

\begin{align}
p(\vo{y}_t  \mid \vo{x}_t  d, \mathcal{M}) &= \mathcal{N} \left (\vo{H}\vo{x_t}, \vo{R} \right)\\
p(\vo{y}_t  \mid d, \mathcal{M}) &= \mathcal{N} \left ( \vo{H} \vo{\mu}_{t|t-1}, \vo{S}_t\right ).
\label{eq:yt_evid}
\end{align}
Recall that $\vo{S}_t = \vo{H} \vo{\Sigma}_{t|t-1} \vo{H}^T + \vo{R}$ and $\vo{S}_t^* = \vo{H}^{*}\vo{\Sigma}_{t|t-1}^*\vo{H}^{*T} + \vo{R}^*$.
Substituting these distributions into Eq. \ref{eq:EGIG_Evidence}, we arrive at
\begin{align}
\text{EGIG}&(\vo{d}, \mathcal{M}, \mathcal{M^*})=\nonumber \\  &\mathbb{E}_{p(\vo{x}_t, \vo{y}_t | \mathcal{M}^*)} \left [ \log \frac{p(\vo{x}_t \mid \vo{y}_t, d, \mathcal{M})}{p(\vo{x}_t \mid d, \mathcal{M})}  \right ] \nonumber\\
=& \frac{1}{2} \bigg( \log \frac{\mid \vo{S}_t \mid}{\mid \vo{R} \mid} - \mathbb{E} [(\vo{y}_t - \vo{H}\vo{x}_t)^T \vo{R}^{-1}(\vo{y}_t - \vo{H}\vo{x}_t)] \nonumber \\
&+ \mathbb{E} [(\vo{y}_t - \vo{H}\vo{\mu}_{t|t-1})^T \vo{S}_t^{-1}(\vo{y}_t - \vo{H}\vo{\mu}_{t|t-1})] \bigg ).
\end{align}
Using Eq. \ref{eq:xy_joint} - \ref{eq:yt_evid}, it is straight forward to compute the means and covariances of $(\vo{y}_t - \vo{H}\vo{x}_t)$ and $(\vo{y}_t - \vo{H}\vo{\mu}_{t|t-1})$,
\begin{align}
(\vo{y}_t - \vo{H}\vo{x}_t) \sim &\mathcal{N}((\vo{H}^* - \vo{H})\vo{\mu}_{t|t-1}^*, \nonumber \\
&(\vo{H}^* - \vo{H})^T \vo{\Sigma}_{t|t-1}^* (\vo{H}^* - \vo{H}) + \vo{R}^*)
\end{align}
\begin{equation}
(\vo{y}_t - \vo{H}\vo{\mu}_{t|t-1}) \sim \mathcal{N}(\vo{H}^* \mu_{t|t-1}^* - \vo{H}\vo{\mu}_{t|t-1}, \vo{S}_t^*).
\end{equation}

Given these distributions, it is useful to define the following variables $\vo{\Delta}_H = \vo{H}^* - \vo{H}$ and $\vo{\Delta}_y = (\vo{H}^* \mu_{t|t-1}^* - \vo{H}\vo{\mu}_{t|t-1})$. Therefore, we can define the EGIG as

\begin{tcolorbox}[ams align]
&\text{EGIG}(\vo{d}, \mathcal{M}, \mathcal{M^*}) = \\
&\frac{1}{2} \bigg( \log \frac{\mid \vo{S}_t \mid}{\mid \vo{R} \mid} - \text{Tr} [\vo{R}^{-1} \vo{\Delta}_H \vo{\Sigma}_{t|t-1}^* \vo{\Delta}_H^T ] -\text{Tr} [\vo{R}^{-1} \vo{R}^*] \nonumber \\
&  + \text{Tr} [\vo{S}_t^{-1} \vo{S}_t^*] -\vo{\mu}_{t|t-1}^{*T} \vo{\Delta}_H^T \vo{R}^{-1} \vo{\Delta}_H \vo{\mu}_{t|t-1}^*\nonumber\\
&+ \vo{\Delta}_y^T \vo{S}_t^{-1} \vo{\Delta}_y\bigg )\nonumber.
\end{tcolorbox}

\noindent \underline{\emph{Infinite Horizon}}: For the infinite horizon case for inferring $\vo{x}_t$ we know that our prior and posterior are Gaussian. Therefore, when computing the EGIG we can use the expression in Eq. \ref{eq:gen_info_gauss} and then compute the expectation over observations $\vo{Y}_t \sim p(\vo{Y}_t | \mathcal{M}^*)$. Here, $r(\vo{X})$ is $p(\vo{x}_t \mid \vo{Y}_t, d, \mathcal{M}^*)$, $p(\vo{X})$ is $p(\vo{x}_t \mid \vo{Y}_t, d, \mathcal{M})$, and $q(\vo{X})$ is $p(\vo{x}_t \mid \mathcal{M})$. By inspection, we see again that the only terms that depend on $\vo{Y}_t$ are the quadratic terms. Therefore, we begin with those terms.

First, we note the asymptotic results: $\vo{\Sigma}_{t|t} \rightarrow \vo{\Sigma}_D$, $\vo{\Sigma}_{t|t}^* \rightarrow \vo{\Sigma}_D^*$, $\vo{\Sigma}_{t|0} \rightarrow \vo{\Sigma}_L$, $\vo{\mu}_{t|0} = \vo{0}$, and $\vo{\mu}_{t|t}^*  \overset{t \rightarrow \infty}{\sim} \mathcal{N} (\vo{0}, \vo{\Sigma}_L^* - \vo{\Sigma}_D^*)$. This gives us that
\begin{align}
\mathbb{E}_{p(\vo{Y}_t | \mathcal{M}^*)} &\left [\left ( \vo{\mu}_{t|t}^* - \vo{\mu}_{t|0} \right )^T \vo{\Sigma_L}^{-1} \left ( \vo{\mu}_{t|t}^* - \vo{\mu}_{t|0} \right ) \right] \nonumber\\
&= \text{Tr} [\vo{\Sigma_L}^{-1} \left (\vo{\Sigma}_L^* - \vo{\Sigma}_D^* \right) ].
\end{align}
For the second expectation, we again rely on results presented in detail in Appendix \ref{sec:asymp_means}. First, $\mathbb{E}_{p(\vo{Y}_t | \mathcal{M}^*)} [\vo{\mu}_{t|t}^* - \vo{\mu}_{t|t}] = \vo{0}$. Second, therefore
\begin{align}
\mathbb{E}_{p(\vo{Y}_t | \mathcal{M}^*)} &\left [\left ( \vo{\mu}_{t|t}^* - \vo{\mu}_{t|t} \right )^T \vo{\Sigma_D}^{-1} \left ( \vo{\mu}_{t|t}^* - \vo{\mu}_{t|t} \right ) \right] \nonumber\\
&= \text{Tr} [\vo{\Sigma_D}^{-1} \vo{M}_\Delta ].
\end{align}
\begin{align}
\vo{M}_\Delta = \text{Cov} (\vo{\mu}_{t|t}^* - \vo{\mu}_{t|t}) = [-\mathbb{I} \;\; \mathbb{I}] \vo{M} [-\mathbb{I} \;\; \mathbb{I}]^T
\end{align}
where $\vo{M}$, the asymptotic covariance matrix of $[\vo{\mu}_{t|t} \;\vo{\mu}_{t|t}^*]^T$, is the solution to the Lyapunov equation given by 
\begin{equation}
    \vo{M} = \vo{\mathcal{A}} \vo{M} \vo{\mathcal{A}^T} + \vo{\mathcal{Q}}
    \label{eq:lypo_m}
\end{equation}
\begin{equation}
\vo{\mathcal{A}}
=
\begin{pmatrix}
\left (\vo{I} - \vo{KH} \right )\vo{A}  & \vo{KH^*A^*} \\
\vo{0} & \vo{A^*}
\end{pmatrix}
\end{equation}
\begin{equation}
\vo{\mathcal{Q}}
=
\begin{pmatrix}
\vo{KS^*K^T}  & \vo{KS^*K^{*T}} \\
\vo{K^*S^*K^T}  & \vo{K^*S^*K^{*T}}
\end{pmatrix}
\end{equation}
Therefore, using these two expectations we arrive at the EGIG for the infinite horizon system,
\begin{tcolorbox}[ams align]
    &\text{EGIG}(\vo{d}, \mathcal{M}, \mathcal{M^*})= \nonumber\\
    &\mathbb{E}_{p(\vo{Y}_t | \mathcal{M}^*)} [\nonumber\\
    &\mathcal{I}_{p(\vo{x}_t \mid \vo{Y}_t, d, \mathcal{M}^*)} [p(\vo{x}_t \mid \vo{Y}_t, d, \mathcal{M}) \mid \mid p(\vo{x}_t \mid \mathcal{M})] ] \rightarrow \nonumber\\
    &\frac{1}{2} \bigg (\text{Tr} [\vo{\Sigma_L}^{-1} \vo{\Sigma}_L^* ]-\text{Tr} [\vo{\Sigma}_D^{-1} (\vo{\Sigma}_D^* + \vo{M}_\Delta )]  + \log \frac{\mid\vo{\Sigma}_L \mid}{\mid\vo{\Sigma}_D \mid} \bigg) \nonumber\\
    &\text{ as } t \rightarrow \infty.
\end{tcolorbox}
\subsection{Expected Discriminatory Information}
While EIG measures efficiency and EGIG measures robustness, we introduce the Expected Discriminatory Information (EDI) criteria to quantify how well an experiment can identify modeling failures. As such, unlike EGIG which is focused on comparing the Bayesian inference solution in the domain of the states $\vo{x}$, EDI compares them in the data domain, $\vo{y}$. Therefore, we can compare models that have different states and forms, e.g. different number of states. The EDI takes inspiration from the use of Bayes factors to compare models. Therefore we define the EDI as the expected Bayes factor given data from a true model $\mathcal{M}^*$

\begin{align}
\text{EDI}(\vo{d}, &\mathcal{M}, \mathcal{M^*}) = \text{D}_{\text{KL}} \left [ p \left (\vo{Y} \mid d, \mathcal{M^*} \right) || p \left (\vo{Y} \mid d, \mathcal{M} \right) \right] \nonumber\\
&=\int p \left (\vo{Y} \mid d, \mathcal{M^*} \right) \log \frac{p \left (\vo{Y} \mid d, \mathcal{M^*} \right)}{p \left (\vo{Y} \mid d, \mathcal{M} \right)} dY.
\end{align}

For the Bayesian filtering context where $\vo{Y}_t = \{y_0 \dots y_t\}$, we can use express the EDI using an iterative update leveraging a similar strategy for computing the model evidence using a Bayesian filter,
\begin{align}
&\text{EDI}(\vo{d}, \mathcal{M}, \mathcal{M^*}, t)\nonumber\\
&=\int p \left (\vo{Y}_t \mid d, \mathcal{M^*} \right) \log \frac{p \left (\vo{Y}_t \mid d, \mathcal{M^*} \right)}{p \left (\vo{Y}_t \mid d, \mathcal{M} \right)} d\vo{Y}_t \nonumber\\
&=\mathbb{E}_{p \left (\vo{y}_t, \vo{Y}_{t-1}  \mid d, \mathcal{M^*} \right)} \left [ \log \frac{p \left (\vo{y}_t, \vo{Y}_{t-1} \mid d, \mathcal{M^*} \right)}{p \left (\vo{y}_t, \vo{Y}_{t-1} \mid d, \mathcal{M} \right)} \right ] \nonumber\\
& = \mathbb{E}_{p \left (\vo{y}_t, \vo{Y}_{t-1}  \mid d, \mathcal{M^*} \right)} \left [\log \frac{p \left (\vo{y}_{t} \mid \vo{Y}_{t-1}, d, \mathcal{M^*} \right)}{p \left (\vo{y}_{t} \mid \vo{Y}_{t-1}, d, \mathcal{M} \right)} \right ] \nonumber \\
&\qquad \qquad+ \mathbb{E}_{p \left (\vo{Y}_{t-1}  \mid d, \mathcal{M^*} \right)} \left [ \log \frac{p \left (\vo{Y}_{t-1} \mid d, \mathcal{M^*} \right)}{p \left (\vo{Y}_{t-1} \mid d, \mathcal{M} \right)} \right] \nonumber \\
 &=\mathbb{E}_{p \left (\vo{Y}_{t-1}  \mid d, \mathcal{M^*} \right)} \Big[\nonumber\\
 &\text{D}_{\text{KL}} \Big[ p \Big(\vo{y}_{t} \mid \vo{Y}_{t-1}, d, \mathcal{M}^* \Big) || p \Big(\vo{y}_{t} \mid \vo{Y}_{t-1}, d, \mathcal{M} \Big) \Big] \Big] \nonumber \\
\label{eq:filter_edi}
\end{align}

Since the EDI is just a KL divergence, for the linear systems we have been studying in this paper it is fairly straight forward to express with the various quantities we have already derived. Therefore, we will state the main results without tenuous algebraic manipulation.\\

\noindent \underline{\emph{Single Step Update}}: For a single time step where the data is generated by the true process model $p(\vo{y_t} \mid d, \mathbb{M}^*)$, (see $\vo{y_t}$ marginal of Eq. \ref{eq:xy_joint}), but we are evaluating $\mathbb{M}$ according to $p(\vo{y_t} \mid d, \mathbb{M})$ (see Eq. \ref{eq:yt_evid}), we can compute the KL divergence for these Guassian distributions using Eq. \ref{eq:KL_MVN}. Giving us

\begin{tcolorbox}[ams align]
    &\text{EDI}(\vo{d}, \mathcal{M}, \mathcal{M^*})= \\
    &\frac{1}{2} \bigg ( \text{Tr} [\vo{S}_t^{-1} \vo{S}_t^*] + \log \frac{\mid \vo{S}_t \mid }{\mid \vo{S}^* \mid} + \vo{\Delta}_{y}^T \vo{S}^{-1} \vo{\Delta}_{y} - s \bigg). \nonumber
\end{tcolorbox}

\noindent $s$ is the number of observations e.g. sensors. Here we recall that $\vo{\Delta}_y = (\vo{H}^* \mu_{t|t-1}^* - \vo{H}\vo{\mu}_{t|t-1})$ and emphasize that $\mu_{t|t-1}^*$ and $\mu_{t|t-1}$ need not be the same dimension since the comparison is happening in the data space.

For the special case were $\vo{H}^* = [\vo{H}, \vo{\Delta}]$, the state of the model $\mathcal{M}^*$ is $\vo{x}_t^* = [\vo{x}_t, \vo{\delta}_t]^T$, $\vo{\mu}_{\delta,t|t-1} = \mathbb{E} [\vo{\delta}_{t|t-1}]$, and $\text{Cov}(\vo{x}_t^*) = \text{Diag}[\vo{\Sigma}_{t|t-1}, \vo{\Gamma}_{t|t-1}]$ e.g. the augmented states is independent of the other states. Then 
\begin{align}
    &\text{EDI}(\vo{d}, \mathcal{M}, \mathcal{M^*})=\frac{1}{2} \bigg ( \text{Tr} [\vo{S}_t^{-1} \vo{\Delta}\vo{\Gamma}_{t|t-1}\vo{\Delta}^T] \\
     &-\log |\mathbb{I} + \vo{S}_t^{-1}\vo{\Delta}\vo{\Gamma}_{t|t-1}\vo{\Delta}^T| + \vo{\mu}_{\delta,t|t-1}^T\vo{\Delta}^T \vo{S}_t^{-1} \vo{\Delta}\vo{\mu}_{\delta,t|t-1} \bigg) \nonumber 
\end{align}
\underline{\emph{Infinite Horizon}}: For the asymptotic case we may choose to ask a slightly different questions when assessing the EDI. Instead of asking about a single $\vo{y}_t$ we can ask about the full trajectory $\vo{Y}_t = \{y_0 \dots y_t\}$. Therefore to compute the EDI, we look to equation \ref{eq:filter_edi}. Under the previous assumptions of asymptotic stability, since we know that the predictives converge and are independent of the observations $Y_t$, we can expect the first term in \ref{eq:filter_edi} to converge to a constant which we call $\Delta_{\text{EDI}}$. Therefore we expect $\text{EDI}(\vo{d}, \mathcal{M}, \mathcal{M^*}, t) \rightarrow t\Delta_{\text{EDI}}$ as $t \rightarrow \infty$ unless $\Delta_{\text{EDI}}=0$ meaning that there is only a finite amount of information to discriminate between the models based on the experiment even in the infinite horizon case. Therefore, $\Delta_{\text{EDI}}$ is the critical quantity for understanding the asymptotic EDI. Using the expression for the Guassian KL divergence Eq. \ref{eq:KL_MVN} and taking the expectation using the asymptotic results found in Appendix \ref{sec:asymp_means}, we find

\begin{tcolorbox}[ams align]
    \Delta_{\text{EDI}}& = \lim_{t\rightarrow \infty} \mathbb{E}_{p \left (\vo{Y}_{t-1}  \mid d, \mathcal{M^*} \right)} \\
    &\ \big[\text{D}_{\text{KL}} [ p \left (\vo{y}_{t} \mid \vo{Y}_{t-1}, d, \mathcal{M}^* \right) || p \left (\vo{y}_{t} \mid \vo{Y}_{t-1}, d, \mathcal{M} \right) ] \big] \nonumber\\
    =&\frac{1}{2} \bigg ( \text{Tr} [\vo{S}^{-1} \vo{S}^*] + \log \frac{\mid \vo{S} \mid }{\mid \vo{S}^* \mid} + \text{Tr} \left [ {\vo{S}^{-1} \vo{M}}_{S} \right ] - s \bigg). \nonumber
\end{tcolorbox}

We recall that $\vo{S}$ and $\vo{S}$ are the stationary predictive covariances for an observation using design $d$ for the models $\mathcal{M}$ and $\mathcal{M^*}$ respectively. The matrix $\vo{M}_{S}$ is given by
\begin{align}
\vo{M}_S &= \text{Cov} (\vo{H}^*\vo{\mu}_{t|t-1}^* - \vo{H} \vo{\mu}_{t|t-1})\nonumber \\
&= [-\vo{HA} \;\;\;\; \vo{H}^*\vo{A}^*] \vo{M} [-\vo{HA} \;\;\;\; \vo{H}^*\vo{A}^*]^T
\end{align}
where $\vo{M}$ is the joint asymptotic covariance matrix of $\vo{\mu}_{t|t}$ and $\vo{\mu}_{t|t}^*$ and is the solution to the previously specified Lyapnuov equation, E.q. \ref{eq:lypo_m}.

\section{Examples}
\subsection{Spring Mass Damper System}
\begin{figure}[ht]
\centering
\vspace{1cm}
\centering
\usetikzlibrary{patterns,decorations.pathmorphing,positioning}
\begin{tikzpicture}[every node/.style={outer sep=0pt},thick,
 mass/.style={draw,thick},
 spring/.style={thick,decorate,decoration={zigzag,pre length=0.3cm,post
 length=0.3cm,segment length=6}},
 ground/.style={fill,pattern=north east lines,draw=none,minimum
 width=0.75cm,minimum height=0.3cm},
 dampic/.pic={\fill[white] (-0.1,-0.3) rectangle (0.3,0.3);
 \draw (-0.3,0.3) -| (0.3,-0.3) -- (-0.3,-0.3);
 \draw[line width=1mm] (-0.1,-0.3) -- (-0.1,0.3);}]

  \node[mass,minimum width=3.5cm,minimum height=2cm,fill=green!50!black] (m1) {$m_1$};
  \node[mass,minimum width=3.5cm,minimum height=2cm,fill=green!50!black,right=1.5cm of
  m1,opacity=.4] (m2) {$m_2$};
  

  \node[left=2cm of m1,ground,minimum width=3mm,minimum height=2.5cm] (g1){};
  \draw (g1.north east) -- (g1.south east);
  \node[right=2cm of m2,ground,minimum width=3mm,minimum height=2.5cm] (g2){};
  \draw (g2.north west) -- (g2.south west);

  \draw[spring] ([yshift=3mm]g1.east) coordinate(aux)
   -- (m1.west|-aux) node[midway,above=1mm]{$k_1$};
  \draw[spring]  (m1.east|-aux) -- (m2.west|-aux) node[midway,above=1mm]{$k_2$};
  \draw[spring]  (m2.east|-aux) -- (g2.west|-aux) node[midway,above=1mm]{$k_a$};

  \draw ([yshift=-3mm]g1.east) coordinate(aux')
   -- (m1.west|-aux') pic[midway]{dampic} node[midway,below=3mm]{$c_1$}
     (m1.east|-aux') -- (m2.west|-aux') pic[midway]{dampic} node[midway,below=3mm]{$c_2$}
     (m2.east|-aux') -- (g2.west|-aux') pic[midway]{dampic} node[midway,below=3mm]{$c_a$};

  \foreach \X in {1,2}  
  {
   \draw[thin,dashed] (m\X.south) -- ++ (0,-1) coordinate[pos=0.85](aux'\X);
   \draw[latex-] (aux'\X) -- ++ (-1,0) node[midway,above]{$x_\X$}
    node[left,ground,minimum height=7mm,minimum width=1mm] (g'\X){};
   \draw[thick] (g'\X.north east) -- (g'\X.south east);
  }

\end{tikzpicture}
\caption{Spring-Mass-Damper System with unknown 2\textsuperscript{nd}mass.}
\label{SMDS}
\end{figure}
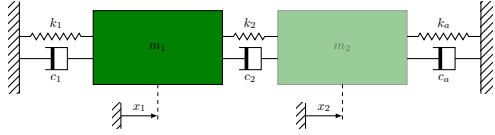
Fig. \ref{SMDS} shows a damped spring-mass system. The equations of motion for this system are
\begin{align}
&{\footnotesize
    \begin{bmatrix}
    1 & 0 & 0 & 0 \\
    0 & 1 & 0 & 0 \\
    0 & 0 & m_1 & 0 \\
    0 & 0 & 0 & m_20
    \end{bmatrix}}
    \frac{d}{dt}
    \begin{bmatrix}
    x_1\\ x_2 \\ v_1 \\ v_2
    \end{bmatrix}
    =\\
    &{\footnotesize\begin{bmatrix}
    0 & 0 & 1 & 0\\
    0 & 0 & 0 & 1\\
    -(k_1+k_2) & k_2 & -(b_1+b_2) & b_2\\
    k_2 & -(k_2+k_3) & b_2 & -(b_2+b_3)
    \end{bmatrix}}\begin{bmatrix}
    x_1\\ x_2 \\ v_1 \\ v_2
    \end{bmatrix}\nonumber
\end{align}
\noindent where $x_1,x_2$ denotes the position of the masses from its rest location. The variables $v_1,v_2$ denoted their linear velocity respectively. The spring constants are $k_1,k_2,k_3$ and the damping coefficients are $b_1,b_2$,$b_3$. This continuous time linear system (CTLS) is then discretized for our analysis.

\begin{figure}[ht]
\centering
\includegraphics[width=.5\textwidth]{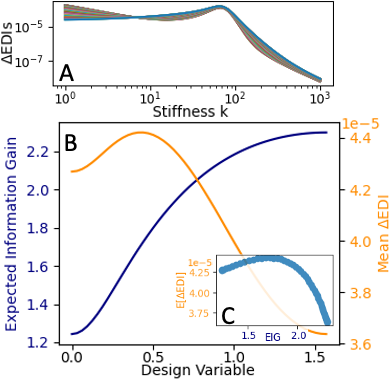}
\caption{Observer design and analysis for the spring mass system. Here, the true model $\mathcal{M}^*$ is a two mass system while the inference model $\mathcal{M}$ is the single mass system. Panel A shows how increasing the stiffness decreases our ability to distinguish between the models. Panels B and C show the trade off between EIG and $\Delta$EDI over our design variable.}
\label{fig:springmass_edi}
\end{figure}

By analyzing the system we can see that under the conditions of high $k_3$ stiffness, low $m_2$ mass, or high $b_3$ damping, that the two mass system should behave closely to a single mass system. Therefore, under these conditions, we would expect the $\Delta$EDI criteria to become small when $\mathcal{M}$ is the one mass system and $\mathcal{M}^*$ is the two mass system. We see panel A in Fig. \ref{fig:springmass_edi} that $\Delta$EDI indeed decreases as we increase the stiffness $k_3$.

We now consider choosing an observer design $d \in [0,\pi/2 ]$ to observe the position and velocity of the known mass, $m_1$, while balancing $\Delta$EDI and EIG. Our, admittedly arbitrary, observer measures the position and velocity of $m_1$ with weights $\cos(d)$ and $\sin(d)$ respectively. The asymptotic EIG objective seeks to maximize information about the position and velocity of $m_1$ according to $\mathcal{M}$. To EDI objective seeks to maximize our ability to asymptotically detect whether $\mathcal{M}$ is plausible versus $\mathcal{M}^*$. Of course we don't know $\mathcal{M}^*$ during the design phase so instead we average $\Delta$EDI over a prior range of stiffnesses from panel A of Fig. \ref{fig:springmass_edi}. We see how EIG varies over the designs as the orange curve in Fig. \ref{fig:springmass_edi}, panel B. While the mean $\Delta$EDI is shown as the navy curve. The trade off between these quantities is show in panel C. Depending on the importance of discrimination vs performance we may choose either only observe the velocity (maximizing EIG) or to sacrifice some EIG to gain better discrimination power by choosing mixed sensor design.

\subsection{F-16 Model}
We use an F-16 aircraft model based from \cite{stevens2015aircraft}\cite{bhattacharya2002nonlinear}. This system originally has 12 states 
\begin{figure}[ht]
\centering
\includegraphics[width=.5\textwidth]{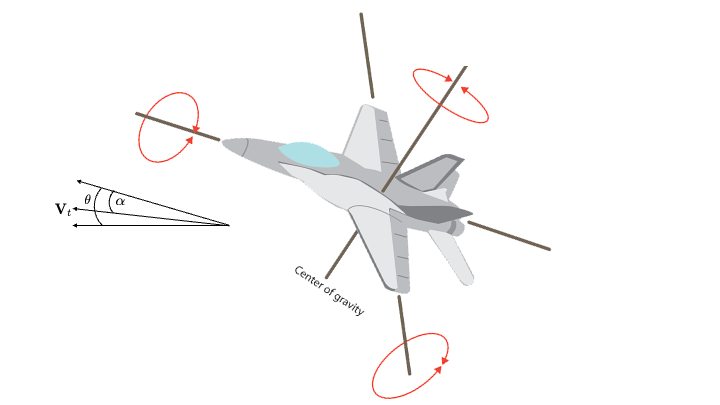}
\caption{F-16 model aircraft with specified states.}
\label{fig:f18}
\end{figure}
 of which we pull out the longitudinal dynamics with states: $\theta$:pitch angle, $V$:velocity, $\alpha$: attack angle, $\dot{\theta}$ and controls: $T$: thurst, $\delta_{ele}$: elevator angle (see Fig. \ref{fig:f18}). We form a reduced-order CTLS using the closed loop system, which is then discretized.

\begin{figure}[ht]
\centering
\includegraphics[width=.48\textwidth]{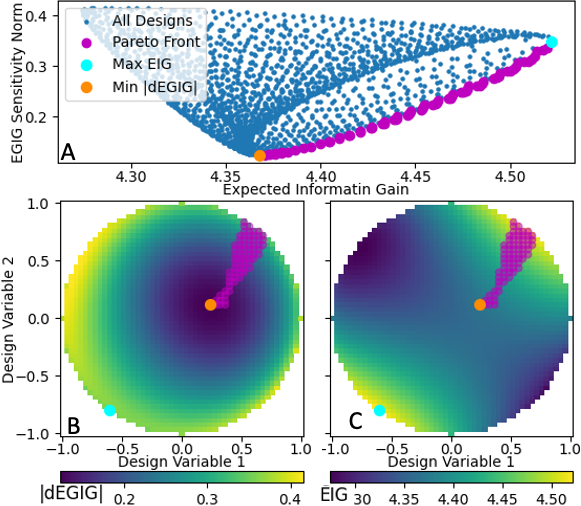}
\caption{Observer design for simplified F-16 model. The true model $\mathcal{M}^*$ is a small perturbation, $\Delta$, in the dynamics from the inference model $\mathcal{M}$. We explore the addition of a new output,  $y_{new} = d_1\theta + d_2\alpha + d_3 \dot{\theta}$ where $d_3$ is constrained by $d_1$ and $d_2$. We measure the improved performance using EIG$(d_1,d_2)$ and robustness using the sensitivity of EGIG e.g., $||\nabla_{\Delta} \text{EGIG}(d_1,d_2)||$. Panel A shows the trade off between these two criteria for different designs. Panels B and C show the projection of these criteria on to the design space.}
\label{fig:f16_design}
\end{figure}

For this model, we seek add an additional output to the observer. This new output has the arbitrary form $y_{new} = d_1\theta + d_2\alpha + d_3 \dot{\theta}$, where $d_1^2+d_2^2+d_3^2 = 1$. When considering these designs, we seek to balance maximizing asymptotic EIG while minimizing asymptotic EGIG. The inference model, $\mathcal{M}$, is the F-16 model with dynamics $\vo{A}$, but the true model, $\mathcal{M}^*$ has dynamics $\vo{A}^* = \vo{A}+\Delta \odot \vo{A}$. So, $\vo{A}^*$ has perturbations scaled relative to $\vo{A}$. Because $\Delta$ is unknown, we instead minimize the sensitivity of EGIG to changes of $\Delta$. Therefore, our metric is the norm, $||\nabla_{\Delta} \text{EGIG}(d_1,d_2)||$. The result is summarized in Fig. \ref{fig:f16_design}. Panel A shows the trade off of different designs between EIG$(d_1,d_2)$ and $||\nabla_{\Delta} \text{EGIG}(d_1,d_2)||$ and the Pareto front of optimal designs (purple). We see that the EGIG is much more sensitive to the design than the EIG, i.e. EGIG varies by about a factor of 4. Therefore, for a robust design we may sacrifice a little asymptotic EIG for meaningful improvement in robustness. Panels B and C show the EIG and EGIG projected on the design space along with the corresponding Pareto set.

We have made the codes to these examples available on GitHub\cite{tommie_nil_github}.

\section{Conclusion}
Maximizing the value of data for inference and prediction requires the careful selection of experimental conditions by modeling the experiment. These models are prone to misspecifications. We propose an information theoretic framework that extends the notion of Expected Information Gain (EIG), typically used in Bayesian experiment design, to address the model mismatch issue. The proposed the Expected Generalized Information Gain (EGIG) captures the information gained or loss with respect to a true model, when the experiment is designed based on a model with discrepancy. On the other hand the proposed Expected Discriminatory Information (EDI) discriminates between models based upon the data generated, which further aids in the model refinement. These three metrics are complementary as EIG emphasizes data efficient experiments, EGIG emphasizes experiments that lead to results that are robust to model discrepancy, and EDI emphasizes experiments that would detect modeling failures. 

\section*{ACKNOWLEDGMENT}
This material is based upon work supported by the U.S. Department of Energy, Office of Science, Office of Advanced Scientific Computing conducted at Sandia National Laboratories. Sandia National Laboratories is a multimission laboratory managed and operated by National Technology and Engineering Solutions of Sandia, LLC., a wholly owned subsidiary of Honeywell International, Inc., for the U.S. Department of Energys National Nuclear Security Administration under contract DE-NA-0003525. This paper describes objective technical results and analysis. Any subjective views or opinions that might be expressed in the paper do not necessarily represent the views of the U.S. Department of Energy or the United States Government. This research used resources of the National Energy Research Scientific Computing Center (NERSC), a U.S. Department of Energy Office of Science User Facility located at Lawrence Berkeley National Laboratory, operated under Contract No. DE-AC02-05CH11231 using NERSC award DOE-ERCAPm3876.

\printbibliography

\section{Appendix}
\subsection{Asymptotic Distribution of Inferred Means}
\label{sec:asymp_means}
Suppose we have a true model of a discrete time, asymptotically stable, linear dynamical system whose variables are denoted with a superscript $*$, while the model used for inference has variables without any superscript. Using a Kalman filter, the inferred mean is then given by
\begin{align}
    \mu_{t} &= \vo{A} \mu_{t-1} + \vo{K} \left (Y_t - \vo{HA}\mu_{t-1} \right) \nonumber\\
    &= \left (\vo{I} - \vo{KH} \right )\vo{A} \mu_{t-1} + \vo{KH^*A^*}\mu_{t-1}^* + \vo{K}\zeta^*.
\end{align}

\noindent here $Y_t \sim \mathcal{N} \left (\vo{H^*A^*}\mu_{t-1}^*, S^*\right)$, so $\zeta^* \sim \mathcal{N} \left (0, S^*\right)$. Similarly, we can define the evolution of mean of the true dynamical system under Kalman filtering as:
\begin{align}
    \mu_{t}^* &= \vo{A^*} \mu_{t-1}^* + \vo{K^*} \left (Y_t - \vo{H^*A^*}\mu_{t-1}^* \right) \nonumber \\
    &= \vo{A^*} \mu_{t-1}^* + \vo{K^*}\zeta^*
\end{align}

First, we note that $E[\mu_t]$ = $E[\mu_t^*]$ = 0, where the expectation is taken over asymptotically long sample trajectories of the true dynamical system.  Second, we note that $E[\mu_{t-1}\zeta^{*T}]=E[\mu_{t-1}^*\zeta^{*T}]=0$ e.g. they are independent. This comes from the fact that only $\mu_{t-1}$ and $\mu_{t-1}^*$ are functions of the trajectory and not $\zeta^*$, or in other words, all the information about the trajectory is captured in the mean estimates. Finally, we note that $\vo{K}$ and $\vo{K^*}$ are known for the asymptotic case by solving the respective DAREs, Eq. \ref{eq:DARE1} - \ref{eq:DARE2}. With that we can express the second moments of $\mu_t$ and $\mu_t^*$ as

\begin{equation}
\vo{M_t}
=
\begin{pmatrix}
E[\mu_t\mu_t^T] & E[\mu_t\mu_t^{*T}] \\
E[\mu_t^*\mu_t^T] & E[\mu_t^*\mu_t^{*T}].
\end{pmatrix}
\end{equation}

In order to solve for the second moments, we define

\begin{equation}
\vo{\mathcal{A}}
=
\begin{pmatrix}
\left (\vo{I} - \vo{KH} \right )\vo{A}  & \vo{KH^*A^*} \\
\vo{0} & \vo{A^*}
\end{pmatrix},
\end{equation}
\begin{equation}
\vo{\mathcal{Q}}
=
\begin{pmatrix}
\vo{KS^*K^T}  & \vo{KS^*K^{*T}} \\
\vo{K^*S^*K^T}  & \vo{K^*S^*K^{*T}}
\end{pmatrix}.
\end{equation}

\noindent Then we can solve for the moments as:

\begin{equation}
    \vo{M_t} = \vo{\mathcal{A}} \vo{M_{t-1}} \vo{\mathcal{A}^T} + \vo{\mathcal{Q}}
\end{equation}

Therefore, for the asymptotic case, we can solve the following Lyapunov equation to find the asymptotic second-order moments, $\vo{M}$:

\begin{equation}
    \vo{M} = \vo{\mathcal{A}} \vo{M} \vo{\mathcal{A}^T} + \vo{\mathcal{Q}}
\end{equation}

\noindent giving us the result that asymptotically:

\begin{equation}
\begin{pmatrix}
\mu_t\\
\mu_t^*
\end{pmatrix}
\sim \mathcal{N} \left (\vo{0}, \vo{M} \right )
\end{equation}

\underline{\emph{Special case $\vo{A}=\vo{A}^*$}}: For the simpler case when we only have one model, the true model, we have the simplified equation given by
\begin{align}
\vo{M^*} = &\vo{A^*} \vo{M^*} \vo{A^{*T}} + \vo{K^*S^*K^{*T}} \nonumber \\
= &\vo{A^*} \vo{M^*} \vo{A^{*T}} + \vo{A^*} \vo{P_D^*} \vo{A^{*T}} + \vo{Q} - \vo{P_D^*} \nonumber\\
\implies &\vo{M^* + P_D^*} = \vo{A^*} ( \vo{M^* + P_D^*} )\vo{A^{*T}} + \vo{Q}
\label{eq:mod_lyp}
\end{align}

\noindent the substitution $\vo{K^*S^*K^{*T}} = \vo{A^*} \vo{P_D^*} \vo{A^{*T}} + \vo{Q} - \vo{P_D^*}$ can be found using the matrix inversion lemma and knowing that $\vo{P_D^*}$ is the solution to Eq. \ref{eq:kalman_filter_last} for the asymptotic case. We observe that Eq. \ref{eq:mod_lyp} is a Lyapunov equation. Thus since we know that $\vo{P_L^*}$ is the solution to the Lyapunov equation for this system. Therefore,

\begin{equation}
\vo{M^*} = \mathbb{E}[\mu_t^*\mu_t^{*T}] = \vo{P_L^*} - \vo{P_D^*}
\label{eq:mu_cov}
\end{equation}
\end{document}